\documentclass[a4]{article}
\newcommand\fverb{\setbox\fverbbox=\hbox\bgroup\verb}
\newcommand\fverbdo{\egroup\medskip\noindent%
			\fbox{\unhbox\fverbbox}\ }
\newcommand\fverbit{\egroup\item[\fbox{\unhbox\fverbbox}]}
\newbox\fverbbox

\newtheorem{theorem}{Theorem}

\def\endprf{\hfill  {\vrule height6pt width6pt depth0pt}\medskip}

\usepackage{amsfonts}
\usepackage{graphicx}
\usepackage{epsfig}

\begin{document}

%Changed 24 June 2010
\title{Infrared behavior of the running coupling in Yang-Mills theory}
%\title{Infrared behavior of the running coupling in scalar field theory}

\author{Marco Frasca \\
%\email[]{marcofrasca@mclink.it}
%\thanks{}
%\affiliation{
Via Erasmo Gattamelata, 3\\ 
00176 Roma (Italy)}


\date{\today}

\maketitle

\begin{abstract}
We compute the Green function of the massless scalar field theory 
% insert abstract hereWe compute the Green function of the massless scalar field theory 
%Added 6 August 2010
in the infrared till the next-to-leading order, providing a fully covariant strong coupling expansion.
Applying Callan-Symanzik equation we obtain the exact running coupling for this case by computing the beta function. This result is applied using a recently proved mapping theorem between a massless scalar field theory and Yang-Mills theory. This beta function gives a running coupling going to zero as $p^4$ in agreement with lattice results presented in Boucaud et al. [JHEP 0304 (2003) 005] and showing that the right definition of the running coupling for a Yang-Mills theory in the infrared is given in a MOM scheme. 
%Added 15 August 2010
The emerging scenario is supporting a quantum field theory based on instantons.
\end{abstract}

%Added 24 June 2010
%\keywords{Nonperturbative Effects, Renormalization Group, Strong Coupling Expansion, Running coupling, Yang-Mills theory}
%\keywords{Nonperturbative Effects, Renormalization Group, Strong Coupling Expansion}

%\dedicated{Dedicated to\ldots\\if you want.}

\section{Introduction}

The study of a quantum field theory in a regime where the coupling becomes increasingly large appears to cope with a lot of difficulties. The problem relies on the absence of useful methods in this regime. This means that the only reliable approach used so far has been lattice computations. 

The model we consider in this paper is a massless scalar field theory. Its relevance relies on the fact that Yang-Mills equations of motion can be reduced to the equation of motion of this theory giving solution for them in the infrared. This permits to obtain the Green function for the gluon \cite{fra1,fra2} and the behavior of the beta function using Callan-Symanzik equations.

For the gluon propagator there is a lot of research activity about, mostly trying to obtain its behavior by solving quantum Yang-Mills theory on the lattice \cite{cuc,ilg1,ste1}. The scenario that is going to emerge shows that the gluon propagator goes to a finite value when the momentum goes to zero. On the contrary, the running coupling goes to zero and, against common wisdom, does not seem to have a fixed point in the infrared.
%Added 24 June 2010
This behavior of the running coupling seems consistent both with the exact beta function in supersymmetric quantum chromodynamics \cite{shif} and the consequent inspired beta function for quantum chromodynamics\cite{sann,boch}. It is interesting to note that, while running coupling of Yang-Mills theory goes to zero at lower energies, nevertheless QCD reaches a fixed point in the same limit due to presence of quarks. This appears a rather counterintuitive result.

A possible definition for the running coupling, the one used in lattice computations, has been given by Alkofer and von Smekal in \cite{as} and is consistent with a recently proposed minimal MOM scheme\cite{ste2}. This definition is quite satisfactory for a qualitative behavior but could be somewhat in disagreement with the real behavior. Indeed, this has been seen in a paper by Boucaud et al. \cite{bou} that show a different definition for the running coupling using a MOM scheme and compute its behavior on the lattice obtaining that this goes to zero as $p^4$. This opens a question about what should be the proper definition for this physical quantity.

An important answer has been given recently through the analysis of experimental data by a series of papers by Prosperi's group \cite{pro1,pro2,pro3}. These authors take the meson spectrum and get the running coupling by a Bethe-Salpeter like formalism. The comparison with analytical methods shows that the experimental data tend to decrease toward zero lying well below an expected fixed point in the infrared.

The current definition of the running coupling, as given in \cite{as}, uses the dressing functions of the gluon and ghost propagator. One has for the gluon propagator
\begin{equation}
    D(p) = \frac{Z(p)}{p^2},
\end{equation}
for the ghost propagator
\begin{equation}
    D^G(p) = \frac{J(p)}{p^2},
\end{equation}
and the running coupling is
\begin{equation}
    \alpha(p) = J^2(p)Z(p).
\end{equation}
We will show consistently that the running coupling as defined in this way is not this one yielding a different behavior with respect to the one discussed in \cite{bou} with a different MOM scheme. The latter goes to zero in the infrared as $p^4$. 
%Added on 24 June 2010
Former definition of the running coupling has been consistently obtained in Ref.\cite{ste2} in a minimal MOM scheme. This appears to go to zero rather as $p^2$. 
%
%Added 25 June 2010
In this paper we analyze the leading order of a strong coupling expansion in a quantum field theory and show how the corresponding two-point function, at the trivial fixed point, indeed satisfies a Callan-Symanzik equation giving a beta function proportional to the coupling, being the proportionality constant the dimensionality of the space-time.
%and consistently describes a trivial theory when the very low energy limit is taken. 
This result appears to be true also for a Yang-Mills theory in the same limit. Both theories display a mass gap in their spectrum. This proves that a set of free particle states exist for these theories to build upon a perturbation theory in a strong coupling limit and to study them in the infrared.
%

%Added 15 August 2010
The propagator at the fixed point behaves in the same way of the one of the Schwinger model of QED in d=1+1, having the generated mass depending on the value of the coupling\cite{sch1,sch2}, and the behavior of the running coupling itself is strongly supporting a scenario based on instantons.

The paper is structured as follows.
%Added 4 August 2010
In sec. \ref{sec11} we discuss the technique we developed in quantum field theory to analyze the limit of a coupling going to infinity and provide a new formulation fully covariant and higher order corrections as well.
In sec. \ref{sec2} we derive the propagator of the massless scalar field theory and show when it solves Yang-Mills equations.
%Added on 24 June 2010
This section is crucial for all this work and we discuss in depth these derivations.
In sec. \ref{sec3} we compare this propagator with lattice results and numerical solution of the Dyson-Schwinger equations. In sec. \ref{sec4} we see show that our propagator solves Callan-Symanzyk equations giving the corresponding beta function. In this way we prove that the running coupling goes to zero in the infrared as $p^4$. Finally, in sec. \ref{sec5} conclusions are given.

%Section added 4 August 2010 
\section{Techniques\label{sec11}}

\subsection{Gradient expansion method}

%Added 4 August 2010
In this section we discuss our approach as devised in \cite{fra3}. This formulation has the disadvantage to be not explicitly covariant and, from a strict mathematical point view, it is not easy to reformulate properly. But the computations give the right answers and, in the end, can be set in the right form. 

%Moved on 4 August 2010
Quantum field theory of a massless scalar field theory is formulated with the partition function
\begin{equation}
   Z[j]=\int [d\phi]\exp\left[i\int d^4x\left(\frac{1}{2}(\partial\phi)^2
   -\frac{\lambda}{4}\phi^4+j\phi\right)\right].
\end{equation}
Our approach is to consider 
%Added 23 June 2010
formally
the limit $\lambda\rightarrow\infty$. This limit has been considered before \cite{kov,pmb,be1,par,be2,be3,coo,be4,svai}. These authors have taken $\frac{1}{2}(\partial\phi)^2$ as a perturbation. This choice has the effect to produce a real singular series needing a proper regularization with no real chance to get finite results in the given limit. The reason for this relies on the fact that some dynamics must be allowed in this limit. This can be obtained by a gradient expansion \cite{fra1,fra2,fra3}. A gradient expansion is obtained by rewriting the above functional as
\begin{equation}
   Z[j]=\int [d\phi]\exp\left[i\int d^4x\left(\frac{1}{2}\dot\phi^2
   -\frac{\lambda}{4}\phi^4+j\phi\right)\right]\exp\left[-i\int d^4x (\nabla\phi)^2\right].
\end{equation}
and considering as a perturbation the gradient part. Now, we note that, in the limit $\lambda\rightarrow\infty$, if a dynamics is allowed, the term $\dot\phi^2$ must be of the same order of the term $\lambda\phi^4$. So, we can safely rescale time as $t\rightarrow\sqrt{\lambda}t$ and rewrite the above functional as
\begin{equation}
   Z[j]=\int [d\phi]\exp\left[i\sqrt{\lambda}\int d^4x\left(\frac{1}{2}\dot\phi^2
   -\frac{1}{4}\phi^4+j\phi\right)\right]\exp\left[-i\frac{1}{\sqrt{\lambda}}\int d^4x (\nabla\phi)^2\right].
\end{equation}
and we see that is just a proof of the fact that the limit $\lambda\rightarrow\infty$ recover a gradient expansion. We need to properly rescale the current $j$ to maintain the right ordering but this is just an arbitrary function. The consequence of this rescaling of time gives us immediately the conclusion that the strong coupling limit is a semiclassical limit \cite{fra4}. So, we take $\phi=\bar\phi + \delta\phi$ being
\begin{equation}
   \ddot{\bar\phi}+\bar\phi^3 = j.
\end{equation}
In the infrared, with the energy going to zero, the following causal approximation does hold \cite{fra6,fra7}
\begin{equation}
   \phi(t)\approx\mu^{-2}\int_0^t dt'G(t-t')j(t')
\end{equation}
being
\begin{equation}
   \ddot G(t)+G(t)^3=\mu^2\delta(t),
\end{equation}
$\mu$ is an arbitrary constant having dimension of energy.
This is a small time approximation and has been recently recovered in a study of nonlinear waves \cite{nlw}. It is somewhat surprising the proper working of Green functions in such nonlinear systems but this turns out useful to maintain all the machinery of quantum field theory. This means that the leading order approximation to hold for our functional in the infrared limit is
\begin{eqnarray}
\label{eq:fgauss}
   Z[j]&\approx& Z[0]\exp{\left[-\frac{i}{2}\int d^4x_1d^4x_2
   \frac{\delta}{\delta j(x_1)}\nabla^2\delta^3(x_1-x_2)\frac{\delta}{\delta j(x_2)}\right]}\times \\ \nonumber
   &&\exp{\left[\frac{i}{2}\int d^4x_1d^4x_2j(x_1)\Delta(x_1-x_2)j(x_2)\right]} 
\end{eqnarray}
being
\begin{equation}
   \Delta(x_1-x_2)=\mu^{-2}\delta^3(x_1-x_2)[\theta(t_1-t_2)G(t_1-t_2)+\theta(t_2-t_1)G(t_2-t_1)]
\end{equation}
and
\begin{equation}
   G(t)=\theta(t)\mu\left(\frac{2}{\lambda}\right)^{\frac{1}{4}}
   {\rm sn}\left[\left(\frac{\lambda}{2}\right)^{\frac{1}{4}}\mu t,i\right]
\end{equation}
%Modified 25 August 2010
being ${\rm sn}$ a Jacobi elliptic function.
%Modified 7 August 2010
%where 
The role of the constant $\mu$ is now clear: This should be considered as a true physical constant of the theory similar to the constant that appears by dimensional transmutation in asymptotic freedom.
%Added 25 August 2010
This can be understood considering the classical equation of motion
\begin{equation}
   \partial^2\phi+\lambda\phi^3=0.
\end{equation}
This admits an exact solution yielded by
\begin{equation}
\label{eq:cls}
   \phi=\mu\left(\frac{2}{\lambda}\right)^\frac{1}{4}{\rm sn}(p\cdot x+\theta,i)
\end{equation}
being $\theta$ a constant phase, provided the dispersion relation
\begin{equation}
   p^2=\mu^2\sqrt{\frac{\lambda}{2}}
\end{equation}
holds. Then, classically we have a free massive solution if $\mu$ is a physical constant having the dimension of energy.
So, we showed that a massless scalar theory in the infrared takes an integrable form.

%Added 4 August 2010
Rather than expanding on this approach to obtain higher order corrections, due to the difficulties to get a fully covariant formulation, we stop this analysis here and we discuss an improved method in the next section.

\subsection{Covariant expansion and higher order corrections}

In order to overcome the difficulties pointed out above about a gradient expansion, we provide here, for the first time, a fully covariant formulation and we are so able to get higher order corrections. It should be pointed out that this strong coupling expansion, differently from other methods, is just dual to standard perturbation theory and it completely shares with it both virtues and defects as we are going to see.

As pointed out in \cite{fra3} and in section above, the possibility to work out a formal expansion in the parameter $1/\lambda$ relies on a rescaling of the time variable $t\rightarrow\sqrt{\lambda}t$. So, a non-trivial expansion is straightforwardly obtained. This kind of approach was devised by us, for standard quantum mechanics and most generally for differential equations, in a series of papers \cite{fraA,fraB,fraC,fraD}. Here we present a covariant form of this expansion and we compute the next-to-leading order correction to the propagator of the scalar field. Our aim is to make clear the behavior of the running coupling of the theory and to obtain at the same time a correction to the propagator to understand the way the spectrum of the theory gets modified.

So, let us consider the action in the generating functional of the scalar field and we take the covariant rescaling $x\rightarrow\sqrt{\lambda}x$. We will get
\begin{equation}
   S=\frac{1}{\lambda}\int d^4x\left[\frac{1}{2}(\partial\phi)^2-\frac{1}{4}\phi^4\right]
   +\frac{1}{\lambda^2}\int d^4xj\phi.
\end{equation}
Now, we take the series
\begin{equation}
   \phi=\phi_0+\frac{1}{\lambda}\phi_1+\frac{1}{\lambda^2}\phi_2+O\left(\frac{1}{\lambda^3}\right)
\end{equation}
and put it into the action, with the rescaling $j\rightarrow j/\lambda$, obtaining
\begin{eqnarray}
S_0&=&\int d^4x\left[\frac{1}{2}(\partial\phi_0)^2-\frac{1}{4}\phi_0^4+j\phi_0\right] \label{eq:ord0} \\
S_1&=&\int d^4x\left[\partial\phi_0\partial\phi_1-\phi_0^3\phi_1+j\phi_1\right] \label{eq:ord1} \\
S_2&=&\int d^4x\left[\frac{1}{2}(\partial\phi_1)^2-\frac{3}{2}\phi_0^2\phi_1^2
+\partial\phi_0\partial\phi_2-\phi_0^3\phi_2+j\phi_2\right]. \label{eq:ord2}
\end{eqnarray}
Now, using the equation of motion
\begin{equation}
\label{eq:phi0}
   \partial^2\phi_0+\phi_0^3=j
\end{equation}
the generating functional at the next-to-leading order takes the simple form
\begin{equation}
   Z[j]\approx e^{i\int d^4x\left[\frac{1}{2}(\partial\phi_0)^2-\frac{\lambda}{4}\phi_0^4+j\phi_0\right]}
   \int[d\phi_1]
   e^{i\frac{1}{\lambda}\int d^4x\left[\frac{1}{2}(\partial\phi_1)^2-\frac{3}{2}\lambda\phi_0^2\phi_1^2\right]}
\end{equation}
after undoing the rescaling in the space-time variables. Now, we can put the first factor of this functional into a Gaussian form plus corrections by solving eq.(\ref{eq:phi0}) noting that $\phi_0$ is a functional of $j$. This can be accomplished generalizing the approach given in \cite{fra6,fra7} to a covariant form. So, let us write the solution of eq.(\ref{eq:phi0}) as
\begin{equation}
    \phi_0=\mu\int d^4x'G(x-x')j(x')+\delta\phi
\end{equation}
being
\begin{equation}
    \partial^2G(x-x')+\lambda[G(x-x')]^3=\frac{1}{\mu}\delta^4(x-x').
\end{equation}
with the constant $\mu$ playing the same role as in the preceding section.
By substitution into eq.(\ref{eq:phi0}) we get the exact equation
\begin{eqnarray}
   \partial^2\delta\phi+\lambda(\delta\phi)^3&=&\mu\lambda\int d^4x'[G(x-x')]^3j(x') \\ \nonumber
   &-&\mu^3\lambda\left[\int d^4x'G(x-x')j(x')\right]^3 \\ \nonumber
   &-&3\mu^2\lambda\delta\phi\left[\int d^4x'G(x-x')j(x')\right]^2 \\ \nonumber
   &-&3\mu\lambda(\delta\phi)^2\int d^4x'G(x-x')j(x').
\end{eqnarray}
We solve this equation by iteration, taking as a first iterate $\delta\phi=0$ and so we get the first correction
\begin{equation}
    \partial^2\delta\phi^{(1)}+\lambda(\delta\phi^{(1)})^3=\mu\lambda\int d^4x'[G(x-x')]^3j(x') 
   -\mu^3\lambda\left[\int d^4x'G(x-x')j(x')\right]^3
\end{equation}
that can be solved with the same approximation for the original equation and we write down
\begin{equation}
    \delta\phi = \mu\lambda\int d^4x'G(x-x')\left\{\mu\int d^4x''[G(x'-x'')]^3j(x'') 
   -\mu^3\left[\int d^4x''G(x'-x'')j(x'')\right]^3\right\}.
\end{equation}
One can check that this approximation reduces to the case of refs.\cite{fra6,fra7} with a small time expansion. So, one has
\begin{eqnarray}
  \phi_0[x;j]&=&\mu\int d^4x'\left\{G(x-x')+\mu\lambda\int d^4x''G(x-x'')[G(x''-x')]^3\right\}j(x') \\ \nonumber
   &-&\mu^4\lambda\int d^4x'G(x-x')\left[\int d^4x''G(x'-x'')j(x'')\right]^3
\end{eqnarray}
from which we are already able to read a first order correction to the propagator. This appears as a functional expansion in powers of $j$. This idea for low-energy QCD, that we exploited here in a rigorous way, was already devised in ref.\cite{rob}. This series is perfectly meaningful in the limit $\lambda\rightarrow\infty$ noting that $G(0)\propto\lambda^{-\frac{1}{2}}$ .

Now, neglecting $O(j^3)$ and higher order terms, we can write down a Gaussian approximation to the generating functional on the lines given in \cite{fra3}. We can write
\begin{equation}
\label{eq:fgen}
   Z[j]\approx e^{\frac{i}{2}\mu\int d^4x'd^4x''j(x')\Delta(x'-x'')j(x'')}
   \int[d\phi_1]e^{i\frac{1}{\lambda}\int d^4x\left\{\frac{1}{2}(\partial\phi_1)^2-\frac{3}{2}\mu^2\lambda
   \left[\int d^4x'\Delta(x-x')j(x')\right]^2\phi_1^2\right\}}
\end{equation}
being
\begin{equation}
\label{eq:pcorr}
  \Delta(x-x')=G(x-x')+\mu^2\lambda\int d^4x''G(x-x'')[G(x''-x')]^3.
\end{equation}
This is a non-trivial approximation that permits us to compute a correction $O(1/\lambda)$ in a strongly coupled quantum field theory. It is evident from this analysis that we are coping with a dual approximation to the standard small perturbation theory sharing identical problems and features. This is a fundamental difference with respect to other approaches. In the next section we complete the computation of the corrections to the propagator.

\section{Two-point functions for massless scalar field and Yang-Mills theory \label{sec2}}

\subsection{Scalar field theory}

%Added 24 June 2010
%We work in 3+1 dimensions as in this case coupling is dimensionless. Analysis can be extended to other dimensions but in this case one needs to introduce a priori an energy scale to work with dimensionless quantities. This constant will then be identified with the energy scale arising as integration constant in the solution of the differential equation for the propagator of the theory.
%

%Added 14 August 2010
In order to compute the propagator of the scalar theory we just need to resum the generating functional (\ref{eq:fgauss}). This is easily accomplished as this functional has the same form of that of a free theory with the mass term exchanged with the Laplacian. So, 
using the series \cite{gr}
\begin{equation}
    {\rm sn}(u,i)=\frac{2\pi}{K(i)}\sum_{n=0}^\infty\frac{(-1)^ne^{-(n+\frac{1}{2})\pi}}{1+e^{-(2n+1)\pi}}
    \sin\left[(2n+1)\frac{\pi u}{2K(i)}\right]
\end{equation}
being $K(i)$ the constant
\begin{equation}
    K(i)=\int_0^{\frac{\pi}{2}}\frac{d\theta}{\sqrt{1+\sin^2\theta}}\approx 1.3111028777.
\end{equation}
%Added 25 June 2010
The functional (\ref{eq:fgauss}) can be immediately evaluated, 
%Added 25 June 2010
evading in this way a problem with a not manifest Lorentz invariance computation proper to a gradient expansion,  
to give
\begin{equation}
   Z[j]\approx Z[0]\exp{\left[\frac{i}{2}\int\frac{d^4p}{(2\pi)^4}j(p)G(p)j(-p)\right]}
\end{equation}
with the full propagator
\begin{equation}
\label{eq:prop}
    G(p)=\sum_{n=0}^\infty\frac{B_n}{p^2-m_n^2+i\epsilon}
\end{equation}
being
\begin{equation}
    B_n=(2n+1)\frac{\pi^2}{K^2(i)}\frac{(-1)^{n}e^{-(n+\frac{1}{2})\pi}}{1+e^{-(2n+1)\pi}},
\end{equation}
and
\begin{equation}
\label{eq:ms}
    m_n = \left(n+\frac{1}{2}\right)\frac{\pi}{K(i)}\left(\frac{\lambda}{2}\right)^{\frac{1}{4}}\mu.
\end{equation}
%Added 25 August 2010
Here the role of $\mu$ is identical to the one of the classical solution (\ref{eq:cls}) and so it must be a physical constant of the theory. Then, we can attach a physical meaning to the spectrum of the theory that has, in this way, a natural cut-off. 

%Added 6 August 2010
We point out that this result is the same in both the approaches devised in the preceding section. In this case, to obtain the result in a closed analytical form, starting from a gradient expansion makes easier to obtain the propagator.

%Added 25 June 2010
We note that this is just the leading order of the strong coupling expansion devised above and, at this order, the theory is trivial as can be seen by the form of the generating functional and further will be shown below with the form of the propagator. So, no renormalization of the coupling takes place at this approximation but the beta function we will get below will give as an insight on the behavior of the coupling at lowering momenta. Renormalization of the coupling is possible to happen going at higher orders in $1/\sqrt{\lambda}$ series, with the bare coupling, much in the same way this happens into the opposite limit $\lambda\rightarrow 0$ and is found in textbooks. 
%Modified on 14 August 2010
%As for today, these higher order terms for this strong coupling expansion have to be evaluated yet but a possible guess is 
We will evaluate a next-to-leading order correction below and we will see
that a propagator like the one obtained by Cornwall could be recovered\cite{corn}.

%Added 24 June 2010
We now prove that a theory with such a propagator is trivial. Let us consider the K\"allen-Lehman representation of the propagator
\begin{equation}
    \Delta(p)=\int_0^\infty d\mu^2\frac{\rho(\mu^2)}{p^2-\mu^2+i\epsilon}.
\end{equation}
For a theory having a discrete spectrum with $N$ excitations we will have 
\[\rho(\mu^2)=\sum_{n=0}^N B_n\delta(\mu^2-\omega_n^2)+\rho_c(\mu^2).\] 
Here $\rho_c(\mu^2)$ will represent the contribution arising from multi-particle states due to interactions between elementary excitations. But if a theory has no interactions this contribution will be zero and we will be left with the propagator (\ref{eq:prop}) in the limit $N\rightarrow\infty$. As we will see below, this should be expected in view of the behavior of the running coupling that is in agreement with a beta function proper to a coupling going to zero at lower momenta.
Indeed, we note the scaling of the propagator with $\lambda^{\frac{1}{4}}$ that already says to us that the coupling should go to zero as $p^4$ in a renormalization group analysis. We will discuss this in sec.\ref{sec4}.

Firstly, we consider the generating functional (\ref{eq:fgen}). This can be evaluated by computing all the functional derivatives and we uncover that it takes the form
\begin{equation}
   Z[j]\approx e^{\frac{i}{2}\mu\int d^4x'd^4x''j(x')\Delta(x'-x'')j(x'')}
   e^{\frac{i}{2}C_4\mu\int d^4x'd^4x''j(x')\Delta(x'-x'')j(x'')}\left[1+O(1/\lambda^2)\right]
\end{equation}
being $C_4$ a constant arising from the propagator of a free massless particle. This can be computed immediately to be $C_4=\frac{3\lambda}{16\pi^2}$ having regularized the integral with the constant $\mu$ as an ultraviolet cut-off.
%This constant is zero by Veltman's integral rule $\int d^Dpp^{-2\nu}=0$. 
%and is obtained by regularizing the value of this propagator at the origin. In d dimensions, one has
%\begin{equation}
%   C_d=\frac{6}{\epsilon(4\pi)^{\frac{d}{2}}}\frac{1}{\lambda\mu^\epsilon}\frac{\mu_{IF}^2}{\mu^2}
%\end{equation}  
%being $\mu_{IF}$ an infrared cut-off taken to go to zero. So, the theory is regularized if we take the limit $\mu_{IF}\rightarrow 0$ before $\epsilon\rightarrow 0$.

Finally, we need to evaluate the first order corrections to the propagator to move from the trivial fixed point in the infrared. So, from eq.(\ref{eq:pcorr}) we have to evaluate
\begin{eqnarray}
\label{eq:gcorr}
   \Delta(p)&=&G(p)+\\ \nonumber
   &&\lambda\frac{1}{\mu^2} G(p)\int\frac{d^4p_1}{(2\pi)^4}\frac{d^4p_2}{(2\pi)^4}
   \sum_{n_1}\frac{B_{n_1}}{p_1^2-m_{n_1}^2}
   \sum_{n_2}\frac{B_{n_2}}{p_2^2-m_{n_2}^2}
   \sum_{n_3}\frac{B_{n_3}}{(p-p_1-p_2)^2-m_{n_3}^2}.
\end{eqnarray}
%The integral on momenta is very well-known in literature \cite{ryd1,ryd2} arising in the two-loop correction to the propagator in weak perturbation theory from the sunset diagram. But here, higher order corrections are all depending on powers of the current. 
We have to evaluate this integral in the limit $\lambda\rightarrow\infty$. Remembering the mass spectrum (\ref{eq:ms})
one gets the following expansion
\begin{equation}
%   G(p)=-\frac{0.5553603669}{\mu^2\lambda^\frac{1}{2}}-\frac{0.2771622868}{\mu^4\lambda}p^2
   G(p)\approx-\frac{0.555}{\mu^2\lambda^\frac{1}{2}}-\frac{0.277}{\mu^4\lambda}p^2
   +O\left(\frac{1}{\lambda^\frac{3}{2}}\right)
\end{equation}
The leading order is the well-known Nambu-Jona-Lasinio limit of a contact interaction that holds for a massive propagator as in our case. So, we see that the limit $\lambda\rightarrow\infty$ is consistent with the low momenta limit $p\rightarrow 0$. Then, the integral is evaluated using again the cut-off $\mu$. This gives finally
\begin{equation}
   \Delta(p)\approx G(p)\left[1-\frac{0.171}{(4\pi)^44\lambda^\frac{1}{2}}-\frac{1.2}{(4\pi)^424
   \lambda}\left(1-\frac{1}{28}\frac{p^2}{\mu^2}\right)+O\left(\frac{1}{\lambda^\frac{3}{2}}\right)\right].
\end{equation}
Indeed, this correction permits us to define a renormalization of the field as
\begin{equation}
   Z_\phi=\sqrt{1-\frac{0.171}{(4\pi)^44\lambda^\frac{1}{2}}+O\left(\frac{1}{\lambda}\right)}\approx
   1-\frac{0.171}{(8\pi)^48\lambda^\frac{1}{2}}+O\left(\frac{1}{\lambda}\right)
   %\approx
   %1-\frac{0.171}{(8\pi)^432\lambda^\frac{1}{2}}-\frac{1.2}{(8\pi)^4192
   %\lambda}\left(1-\frac{1}{28}\frac{p^2}{\mu^2}\right)+O\left(\frac{1}{\lambda^\frac{3}{2}}\right)
\end{equation}
and correspondingly to the coupling as $\lambda_R=Z_\phi^2\lambda$. In this way we can conclude that, at the next-to-leading order, the spectrum of the theory does not change and the theory appears renormalizable. We emphasize that this situation makes our approach exactly dual to standard weak perturbation theory and meaningful conclusions can be drawn.

%This result is the Cornwall's correction \cite{corn} and agrees with Schwinger's idea that the propagator has a pole at $p=0$, arising dynamically, due to higher order corrections \cite{sch1,sch2}. We emphasize that, in our case, the cut-off $\mu$ is physically motivated and measured experimentally.

%and is obtained by regularizing the value of this propagator at the origin. In d dimensions, one has
%\begin{equation}
%   C_d=\frac{6}{\epsilon(4\pi)^{\frac{d}{2}}}\frac{1}{\lambda\mu^\epsilon}\frac{\mu_{IF}^2}{\mu^2}
%\end{equation}  
%being $\mu_{IF}$ an infrared cut-off taken to go to zero. So, the theory is regularized if we take the limit $\mu_{IF}\rightarrow 0$ before $\epsilon\rightarrow 0$.
%and we compute the 2-point function as
%\begin{equation}
%   \bar\Delta(x-x')=\left.\frac{1}{Z}\frac{\delta^2Z}{\delta j(x)\delta j(x')}\right|_{j=0}
%\end{equation}
%and the 4-point function as
%\begin{equation}
%   \bar\Delta^{(4)}(x,x',x'',x''')=\left.\frac{1}{Z}\frac{\delta^4Z}{\delta j(x)
%   \delta j(x')\delta j(x'')\delta j(x''')}\right|_{j=0}.
%\end{equation}
%So, 
%for the 2-point function 
%we get 
%\begin{equation}
%   \bar\Delta(p)=\Delta(p)\left[1-\frac{\lambda\mu^{\epsilon+2}}{16\pi^2}\Delta(p)+\ldots\right]
%\end{equation}
%where we used the natural cut-off of the theory $\mu$ that fixes the energy scale for the theory to hold. Here the singularity arises from a free-particle propagator but the limit $\lambda\rightarrow\infty$ is finite as can be easily checked. This is a 2-loop correction that will modify $\mu$ making it depending on powers of $1/\sqrt{\lambda}$.

\subsection{Yang-Mills theory}

This approach is rather general and gives a theoretical framework to treat a quantum field theory in the infrared limit. So, an important step is its application to a pure Yang-Mills theory. As already said, there is a lot of activity about the solution of this theory in the infrared mostly because there is a serious interpretation problem for the spectrum of the light unflavored mesons. For a Yang-Mills theory we will have to solve the equation \cite{ps}
\begin{equation}
\label{eq:ym}
    \partial^\mu\partial_\mu A^a_\nu-\partial_\nu(\partial^\mu A^a_\mu)
    +gf^{abc}A^{b\mu}(\partial_\mu A^c_\nu-\partial_\nu A^c_\mu)+gf^{abc}\partial^\mu(A^b_\mu A^c_\nu)
    +g^2f^{abc}f^{cde}A^{b\mu}A^d_\mu A^e_\nu=j_\nu^a.
\end{equation}
being $f^{abc}$ the structure constants of the Lie group and $g$ the coupling constant. When we try to solve for a gradient expansion these equations we meet a severe problem. Most solutions are just chaotic \cite{sav1,sav2,sav3} and are useless to build a quantum field theory. In order to get a set of solutions to start building a quantum field theory, we can apply a mapping theorem recently proved\cite{fras1,fras2,fras3}
\begin{theorem}[Mapping]
\label{teo1}
An extremum of the action
\begin{equation}
    S = \int d^4x\left[\frac{1}{2}(\partial\phi)^2-\frac{\lambda}{4}\phi^4\right]
\end{equation}
is also an extremum of the SU(N) Yang-Mills Lagrangian when one properly chooses $A_\mu^a$ with some components being zero and all others being equal, and $\lambda=Ng^2$, being $g$ the coupling constant of the Yang-Mills field, when only time dependence is retained. In the most general case the following mapping holds
\begin{equation}
    A_\mu^a(x)=\eta_\mu^a\phi(x)+O(1/\sqrt{N}g)
\end{equation}
being $\eta_\mu^a$ constant, that becomes exact for the Lorenz gauge.
\end{theorem}
Recent lattice computation in 2+1 dimensions have given strong support to this theorem \cite{fri}.
So, with this choice, the equation for the Green function of a 
%Added 25 June 2010
classical 
Yang-Mills theory, in a strong coupling limit, reduces to the one of the case of the massless scalar theory with the substitution $\lambda\rightarrow Ng^2$ for a SU(N) group and this is just the 't Hooft coupling as should be expected \cite{fra2,fras1}. So, finally, e.g. the gluon propagator in the Landau gauge can be written as
\begin{equation}
    D^{ab}_{\mu\nu}(p)=\delta^{ab}\left(\eta_{\mu\nu}-\frac{p_\mu p_\nu}{p^2}\right)G(p)+O(1/\sqrt{N}g)
\end{equation}
being $G(p)$ given in eq.(\ref{eq:prop}) with $\lambda$ substituted by $Ng^2$. This means that we are approximating the solution of eq.(\ref{eq:ym}) with
\begin{equation}
    A^a_\mu(x)\approx\int d^4x'D^{ab}_{\mu\nu}(x-x')j^{b\nu}(x')+O(1/\sqrt{N}g).
\end{equation}
This latter equation should be checked on lattice computations.

It would not be difficult to prove that, with this class of solutions of Yang-Mills equations, the ghost propagator is the one of a free particle as it decouples\cite{fras1}:
\begin{equation}
    D^G(p)=\frac{1}{p^2}+O(1/\sqrt{N}g).
\end{equation}
giving in our case $J(p)=1$ for its dressing function. We can similarly get for the dressing function of the gluon propagator, $Z(p)=G(p)p^2$. We see that when $p$ goes to zero then we have
\begin{equation}
    Z(p)\approx G(0)p^2
\end{equation}
with $G(0)\ne 0$ as can be immediately realized. Then we get the infrared indexes $k_1=0$ for the ghost and $k_2=1$ for the gluon clearly different from those given in \cite{as,ilg}. So one gets for the running coupling as defined by Alkofer and von Smekal $\alpha(p)=G(0)p^2$ that goes to zero in the infrared limit $p\rightarrow 0$. We then get a third index $k_3=2$ for the running coupling. The value of this third index will be the main point of our discussion in sec. \ref{sec4}.

%Added 25 June 2010
These arguments based on the mapping theorem can be kept just at a classical level. We will not do quantum field theory for Yang-Mills theory here but we will compare the propagator for the Yang-Mills field with numerical results in the next section. This will show that the solutions selected with the mapping theorem are indeed the one that determine the infrared behavior of Yang-Mills field.

\section{Numerical results \label{sec3}}
Having obtained an explicit expression for the gluon propagator we have to see how this expression compares with respect to numerical results. We will consider two kind of comparison for our aims: firstly we compare it with the most advanced lattice results and secondly we will consider a numerical solution of the Dyson-Schwinger equation.

The problem with Dyson-Schwinger equation is that a scenario has been proposed \cite{as} enforcing the view that the gluon propagator should go to zero at small momenta, the ghost propagator should go to infinity faster than the free particle case and that the running coupling reaches a fixed point in the same limit. Recent data on lattice are showing that this view
%Modified 14 August 2010 
%is
seems
not correct \cite{cuc,ilg1,ste1}. Besides, data extracted from meson spectrum also showed that the running coupling is not going to reach a fixed point bending clearly toward zero \cite{pro1,pro2,pro3}.

About numerical solutions of Dyson-Schwinger equation there exist two kind of solutions: on a compact manifold \cite{cs1,cs2} and D=3+1 \cite{an}. For the solution on a compact manifold there are contradictory results but it is clearly seen the running coupling bending toward zero. We cannot use these data for comparison
%Added 25 June 2010
as we have no mathematical proof that our formulas can be used for this case
. But we can compare with the results due to Aguilar and Natale \cite{an}. Solving Dyson-Schwinger equations in D=3+1 these authors were able to recover the scenario that is presently emerging from lattice computation. They get a gluon propagator going to a finite value at lower momenta, the ghost propagator converging to the free one and the running coupling as defined by Alkofer and von Smekal going to zero. Then, we must consider these as the proper reference data to compare.

An important point to be noticed is that there is only one parameter to be computed to compare our propagator with numerical data and this is the gluon mass that we write explicitly as $m_g=(\pi/2K(i))\Lambda(Ng^2/2)^{\frac{1}{4}}$. From this value one gets the integration constant $\Lambda$ that is a number to be decided experimentally. 
%Added 25 June 2010
This number must coincide with the same constant appearing in asymptotic freedom calculations by dimensional transmutation\cite{fram}. We just note that some authors proposed a dependence on the energy scale for this number\cite{bou3}. Similarly, to get a satisfactory agreement with lattice computations of Yang-Mills spectrum \cite{tep1,tep2,morn}, one needs to identify $\sqrt{\sigma}=\Lambda(Ng^2/2)^{\frac{1}{4}}$ with the square root of the string tension of QCD generally taken to be around 400 MeV.
%Added on 14 August 2010
%This spectrum should be modified in view of eqs.(\ref{eq:1corr}) and (\ref{eq:mur}). In QCD this turns into a modification of the string tension that will be also depending slightly on momenta at low energies as also expected in \cite{bou3}.
%

\begin{figure}[tbp]
\begin{center}
\includegraphics[width=320pt]{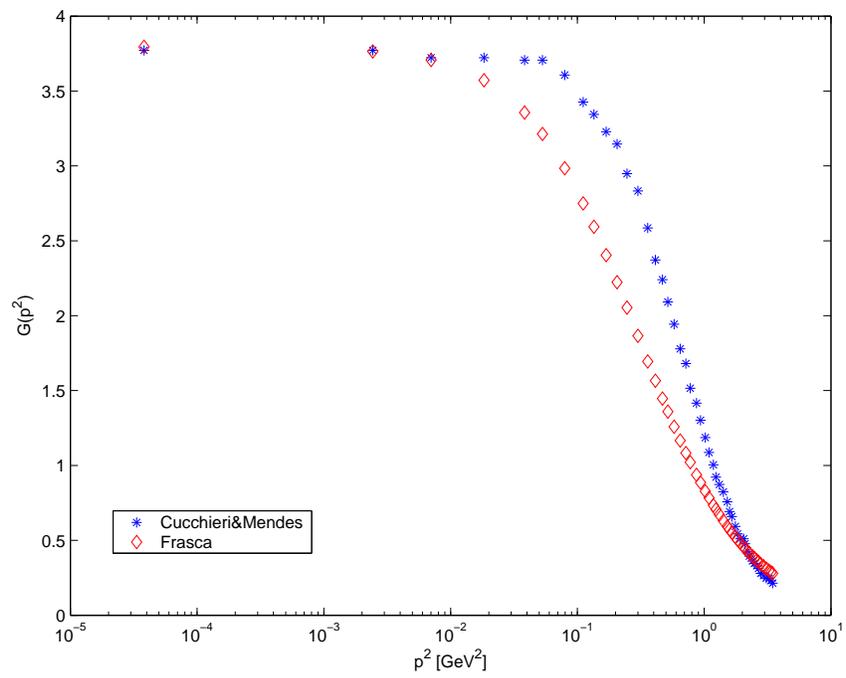}
\caption{\label{fig:fig1} Gluon propagator compared to lattice data in \cite{cuc}.}
\end{center}
\end{figure}

Firstly, let us see as our propagator compare to the most up to date lattice data \cite{cuc}. We see the results in fig.\ref{fig:fig1}. At lower momenta the agreement is perfect with a gluon mass of 545 MeV. There are volume effects but these are seen only in the intermediate energy region. This was also seen in \cite{fra1}.
It should be expected that increasing the lattice volume should make both curves coincide.
%\FIGURE{
%}

\begin{figure}[tbp]
\begin{center}
\includegraphics[width=320pt]{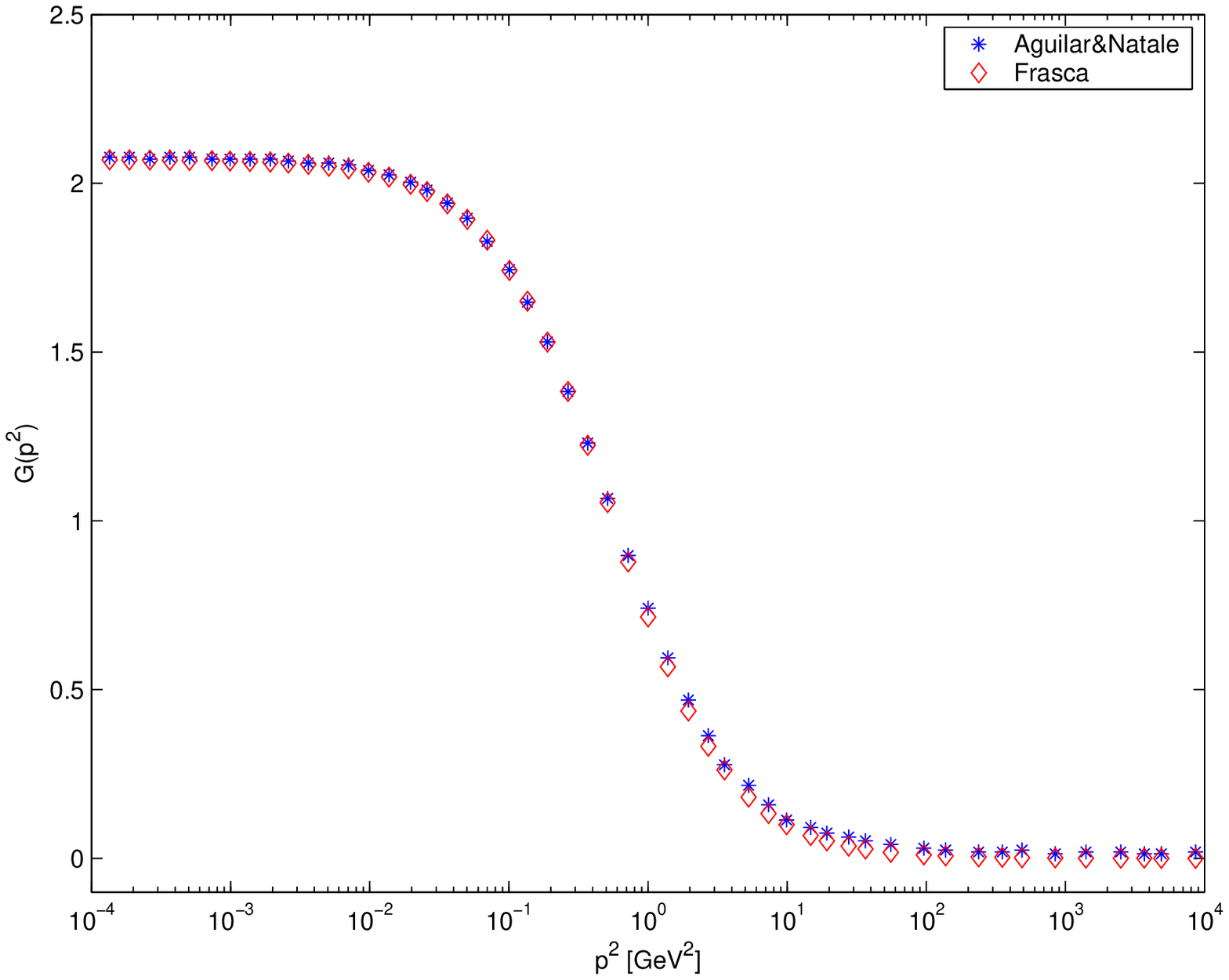}
\caption{\label{fig:fig2} Gluon propagator compared to numerical data in \cite{an}.}
\end{center}
\end{figure}

Finally, we compare our data with the numerical results of Aguilar and Natale \cite{an}. In this case we must have complete coincidence and this is exactly what happens. This can be seen in fig.\ref{fig:fig2}. The fit is obtained with a gluon mass of 738 MeV.
%\FIGURE{
%}
From this figure is clearly seen that Aguilar and Natale hit numerically our propagator. With a proper fit they could have obtained all the full glueball spectrum.

The agreements with lattice data and numerical solution are astonishingly good. 
%Added 25 June 2010
%So, the next step is to draw one more important consequence from our solution.
We can draw a relevant conclusion from this: Numerical studies of Yang-Mills theory show the soundness of the mapping theorem in selecting a set of solutions that are adequate to build a quantum field theory in the infrared limit.
 
\section{Callan-Symanzyk equation and running coupling \label{sec4}}

A proper definition of the running coupling in the infrared is not a trivial matter. An interesting analysis has been presented in \cite{pro1}. Lattice computations use the generally accepted definition due to Alkofer and von Smekal \cite{as}. It is an obvious matter that a proper definiton can be only obtained having the gluon propagator and solving the Callan-Symanzyk equation. This is exactly our aim. Let us point out that in \cite{bou} an analysis was carried out through lattice computation, assuming a MOM scheme to hold for the running coupling in the form
\begin{equation}
\alpha_{\rm  MOM}(p)  \equiv \frac {g_R^2}{4\pi} = \frac 1{4\pi}
\left[\frac{G^{(3)}(p^2,p^2,p^2)}{(G^{(2)}(p^2))^3}(p^2G^{(2)}(p^2))^{3/2}\right]^2
\end{equation}
being $G^{(3)}(p^2,p^2,p^2)$ the three-gluon Green function and $G^{(2)}(p^2)$ the two-point function corresponding to our $G(p)$. These authors prove that, on the lattice, 
$\alpha_{\rm  MOM}(p)\propto p^4$. We see that, if $G^{(2)}(0)=constant$, to have the running coupling going as
$p^4$ implies $G^{(3)}(p^2,p^2,p^2)\propto 1/p$ and also this is in agreement with Boucaud et al. conclusions.
Indeed, these authors have given a description of the infrared behavior of a pure Yang-Mills theory
in close agreement with the one that is presently emerging from lattice computations (see e.g. \cite{bou2}
and refs. therein).

%Added 14 August 2010
The propagator given in eq.(\ref{eq:prop}) is the propagator of the theory at the trivial infrared fixed point. But it must satisfy Callan-Symanzik equation. This leading order propagator is proper to a massive theory notwithstanding we started from a massless theory and, much in the same way of the Schwinger model in d=1+1, the mass arises from the value of the coupling that, at the next-to-leading order would require just a rescaling. The presence of the coupling in the propagator at the trivial infrared fixed point and the need to satisfy Callan-Symanzik equation forces the theory to provide the behavior of the running coupling. This is what we are going to get.

The propagator of the massless scalar theory we have derived above must satisfy the Callan-Symanzik equation
\begin{equation}
   \mu\frac{\partial G({\bf x},t)}{\partial\mu}-\beta(\lambda)\frac{\partial G({\bf x},t)}{\partial\lambda}
   +2\gamma G({\bf x},t)=0
\end{equation}
%Modified 24 June 2010
and the minus sign arises from the fact that we used the physical scale $\mu$ rather than the running scale being their ratio the physical quantity of interest\cite{ps}.
%and 
Setting ${\bf x}=0$ we have immediately
\begin{equation}
   \mu\frac{\partial G(t)}{\partial\mu}-\beta(\lambda)\frac{\partial G(t)}{\partial\lambda}+2\gamma G(t)=0
\end{equation}
or, working with momentum, one has
\begin{equation}
    \mu\frac{\partial G(p)}{\partial\mu}-\beta(\lambda)\frac{\partial G(p)}{\partial\lambda}+2\gamma G(p)=0.
\end{equation}
We see immediately that this is true if
\begin{equation}
   \beta(\lambda) = 4\lambda
\end{equation}
and
\begin{equation}
   \gamma = -1
\end{equation}
giving us the behavior of the running coupling we looked for. This result is in agreement with recent analysis \cite{sus1,sus2} where such a form of beta function in the strong coupling limit was postulated. Similarly, from AdS/CFT correspondence a similar conclusion was drawn \cite{pod}. 

So, this theory can be considered trivial in the sense of Wilson \cite{wk}. Indeed, we know that
\begin{equation}
   \beta(\lambda)=\mu\frac{d\lambda}{d\mu}
\end{equation}
giving immediately
\begin{equation}
\label{eq:rc}
   \lambda(p)=\lambda_0\frac{p^4}{\mu^4}.
\end{equation}
Now, with the substitution $\lambda\rightarrow Ng^2$ that maps the scalar theory on the Yang-Mills theory 
we get immediately the result given in \cite{bou}. This permits us to draw immediately the conclusion that the definition $\alpha_{MOM}(p^2)$ is the proper one for the running coupling in the infrared. This is expected to go to zero with a fourth power law reaching no fixed point. This gives index $k_3$ being 4. Such a result can be understood if we recall what we have said above about the ghost propagator. The ghost in the infrared limit decouples from the gluon and then, a definition of the running coupling implying the ghost propagator has no physical meaning.
%Added 25 June 2010
Finally, we note how this definition of the running coupling makes all our argument fully consistent: Eq.(\ref{eq:rc}) implies that the coupling goes to zero, i.e. no interaction, as momenta decrease. 

Similarly, we have
\begin{equation}
   \gamma = \frac{1}{2}\frac{\mu}{Z}\frac{dZ}{d\mu}
\end{equation}
being $Z$ the renormalization constant of the field. We see immediately that $Z=p^2/\mu^2$ being this the expected scaling for the field at the trivial fixed point. We just note that in the infrared limit particles are not gluons but rather the particles in the mass spectrum of the theory that should be properly called {\sl glueballs}.

\section{Conclusions\label{sec5}}

%Added on 14 August 2010
We emphasize that, differently from the techniques presented in \cite{kov,pmb,be1,par,be2,be3,coo,be4,svai}, our approach is exactly dual to the standard weak perturbation theory sharing with it all the advantages and defects. We just note that, at the one-loop computation we need renormalization of the wave-function and the coupling. Indeed, we provided a consistent framework that, in principle, should permit computations to any desired order.

The results we obtained, mostly due to the similarity with the Schwinger model and the behavior of the running coupling in agreement with a liquid of instantons, should put the basis for an understanding of them through instantons also in virtue of their arising from peculiar classical solutions both of the scalar theory and Yang-Mills theory mapping each other.

The conclusions to be drawn describe a scenario completely different from the one discussed in literature in these latter years. This is due mostly to lattice data that have shacked during these years the common beliefs that were going to form on known theoretical methods.

Presently, evidence is mounting that the gluon propagator reaches a finite value for momenta going to zero. In this paper we have shown that the proper definition of the running coupling in the infrared should be taken in a MOM scheme. This is shown to go to zero, not to a fixed point, with a fourth power of momentum. Finally, this result agrees well with the view that the ghost in the infrared decouples from the gluon and behaves as a free particle.

As this scenario is emerging from numerical solutions of the Yang-Mills quantum field theory and from phenomenological analysis, increasing confirmations in the future years have to be expected.

\end{document}